\documentstyle[11pt,newpasp,twoside,epsf]{article}
\def\edcomment#1{\iffalse\marginpar{\raggedright\sl#1\/}\else\relax\fi}
\reversemarginpar

\begin{document}
\title{Disruption of satellites in cosmological haloes}
\author{Giuliano Taffoni}
\affil{SISSA, via Beirut 2-4, 34014 Trieste -- Italy}
\author{Lucio Mayer}
\affil{Department of Astronomy, University of Washington,
Seattle, USA}
\author{Monica Colpi}
\affil{Dipartimento di Fisica, Universit\`a Degli Studi di
Milano Bicocca, Piazza della Scienza 3, I -- 20126 Milano, Italy}
\author{Fabio Governato}
\affil{ Osservatorio Astronomico di Brera,
via Bianchi 46, I--23807 Merate (LC) - Italy}

\begin{abstract}
  We investigate how the survival of dark matter satellites
  inside virialized halos depends on tidal stripping and
  dynamical friction. We use an analytic approach and then  
  compare the results with  N-Body simulations. 
  Both the satellites and the primary halos are similar
  to cosmological haloes and have NFW density profiles. 
  Satellites can either merge with 
  the primary halo or continue to move on 
  barely perturbed orbits, eventually being disrupted, depending on the 
  relative strength of friction and tidal forces. 
  We propose that their actual fate depends simply on their mass 
  ratio relative to the primary halo.
\end{abstract}

\section{Introduction}
In the current view, galaxy formation occurs within dark matter haloes
while these grow and evolve through a complex hierarchy of mergers.
Understanding the dynamical evolution of haloes is thus essential in
order to understand the evolution of galaxies. High-resolution N-body
simulations show that small haloes, once they are accreted by a larger
halo, can retain their identity and become substructures (Moore, Katz
\& Lake 1996; Tormen 1997; Ghigna et al. 1998).  These substructures
are affected by various dynamical mechanisms which can lead to their
disruption. The dynamical friction drives the satellites towards the
center of the primary system where they can ultimately merge. The
tidal forces exerted by the primary halo cause the satellite to lose
mass and can lead to its evaporation (Gnedin \& Ostriker 1997; Gnedin,
Hernquist \& Ostriker 1999; Taylor \& Babul 2000; Taffoni et al. 2001
in prep).

As it has been already pointed out in Colpi, Mayer \& Governato
(1999), a satellite can be disrupted before friction has eroded
significantly its orbit. In semianalytical models that attempt to
trace the evolution of galaxies within haloes (i.e.  Kauffmann et al.
1999; Somerville \& Primack 1999; Cole et al. 2001), a merging event
between two or more haloes corresponds to the complete loss of their
identity and the galaxies within them are evolved as if they were
detached from their original haloes. In particular, the possibility
that one of them might be disrupted along with its halo is completely
neglected by semianalytical models.  Tidal disruption is instead known
to be likely for other stellar systems subject to a strong tidal
field, like the globular clusters in our Galaxy (Gnedin \& Ostriker
1997).  Clearly, the mechanism of disruption can be studied properly
only taking into account the simultaneous effect of dynamical friction
and tidal stripping.

\section{The dynamical  evolution in a NFW halo}

The full dynamical evolution of the satellites must be studied for haloes 
analogous to those forming  in cosmological simulations, here described 
by the  so called NFW density profile  (Navarro, Frenk \& White 1996):
\begin{equation}
\label{eq:rho}
\rho(r)= {M_{\rm v} \over 4 \pi R_{\rm v}^3}  {\delta_{\rm c} \over 
c \, x  (1+c\,x)^2} \;,
\end{equation}
where $x=r/R_{\rm v}$, $R_{\rm v}$ is the virial radius, $M_{\rm v}$
is the mass of the halo inside the virial radius and $\delta_{\rm
  c}=c^3/(\log(1+c)-c/(1+c))$ with $c=r_{\rm s}/R_{\rm v}$ ($r_{\rm
  s}$ is a scale radius).

In a spherically symmetric system the orbit is planar and can be
determined using the planar polar coordinates $r(t)$ and $\theta(t)$
solving the equation of motion for the static NFW spherical potential
$\phi$ (see e.g. Binney \& Tremain 1987).  The motion of a satellite
is then determined by the initial angular momentum $J$ and the orbital
energy $E$ which can be expressed in terms of the radius of a circular
orbit with the same energy as the considered one $r_{\rm c}(E)$ and
the circularity $\epsilon=J/J_{\rm c}$ ($J_{\rm c}=V_{\rm c}\cdot
r_{\rm c}$).

\subsection{Dynamical friction}

During its motion, the satellite induces a perturbation on the density
field of the primary halo; the net result of this distortion is a
back-reaction force that decelerates the satellite driving it towards
the center of mass of the main halo.  In the limit of a uniform
infinite collisionless background Chandrasekhar (1943) developed a
simple theory to model this {\em dynamical friction force.}  A body of
mass $M_{\rm s}$ moving with velocity $\bf {v}$ relative to a
background of stars with mass $m\ll M_s$ and density $\rho,$
experiences a drag force
\begin{equation}
{\bf {f}}_{df}=- 4 \pi G^2 M_{\rm s}^2\rho\;\log(\Lambda)
{ \Xi(v/(\sqrt 2\sigma)) \over
v^3} \bf {v} \;,
\end{equation}
where $\sigma$ is the one-dimension velocity dispersion of the stars,
and $\Xi(x)={\mathrm erf}(x)-2x/\sqrt \pi\; \exp(-x^2)$ (see e.g.
Binney \& Tremain 1987).  The normalization of the force is given by
the so called {\em Coulomb logarithm} $\log \Lambda$. This is usually
set as $\log \Lambda=\log (M_{\rm s}/M_{\rm halo})$ (Binney \& Tremain
1987; Lacey \& Cole 1993; provided $M_{\rm s}/M_{\rm halo}\ll 1$).
We treat the frictional drag on the satellite as local and consider
$\rho(r)$ and $\sigma(r)$ as described by NFW.  Dynamical friction in
a non-uniform self-gravitating stellar background with NFW density
profile is treated self-consistently using the theory of linear
response (Colpi \& Pallavicini 1998) in Taffoni et al. (2001).

\subsection{Tidal effects}
The overall effect of the tidal perturbation is the progressive
evaporation of a satellite; this process takes place during the entire
orbital evolution and it is generally sensitive both to the internal
properties of the satellite and of the surrounding halo.  We
distinguish two tidal effects: a tidal truncation ({\em tidal cut}),
originated by a steady tidal force, and a secular evaporation effect
({\em tidal shock}), induced by repeated gravitational shocks which
take place internal to the satellite at each pericentric passage.

A satellite orbiting in a halo is tidally truncated at its tidal
radius $r_{\rm t}$ which is defined as the distance of the center of
mass of the satellite from the saddle point of the potential of the
total system.  Loosely speaking it corresponds to the radius at which
the mean density of the satellite is of the order of the mean density
of the halo within $R$ (the distance of the two centers of mass):
$\bar\rho_{\rm s}(r_{\rm t}) \approx \bar \rho_{\rm halo}(R)$.  Such radius
will be a function of time if the orbit is perturbed by, e.g.,
dynamical friction.

At each periastron passage the satellite crosses very rapidly the
central and more concentrated regions of the primary halo where tides
are strongest, providing internal heating.  Such kind of interactions
are called tidal shocks (Spitzer 1978) and usually last a time small
compared to the mean internal dynamical time of the satellite.

We will use the results derived by Gnedin, Hernquist, \& Ostriker
(1999) to describe the amount of heating due to tidal shocks.  During
an orbital period $T$ the tidal force produces a global variation on
the velocity of the internal fluid.  This velocity change causes a
reduction of the binding energy of the system, which is quadratic in
the perturbation $\langle \Delta E \rangle= (1/2)\langle \Delta
v^2\rangle.$ An important role is played by the {\em adiabatic
  corrections}: Gnedin, Lee \& Ostriker (1998) (see also Weinberg
1994) showed that the effect of the tidal shocks can be properly
modeled by multiplying the impulsive energy change $\Delta E$ for a
correction which takes into account that the response of the stars can
be adiabatic to a certain degree (i.e. the system can partially
readjust to the time-dependent tidal field).  The actual energy
associated with the shock is reduced and reads:
\begin{equation}
\langle \Delta E \rangle_{\rm act}=
\langle \Delta E \rangle A(\omega\tau)\;,
\end{equation}
where $A(\omega \tau)$ is a given function of the orbital frequency of
stars $\omega(r)$ relative to the shock duration time $\tau.$

At every periastron passage the satellites reduces its binding energy
by an amount $\langle \Delta E_{\rm hm} \rangle_{\rm act}$ which is
evaluated at the half mass radius of the tidally truncated satellite.
As a consequence of this energy gain, the satellite expands and,
eventually, evaporates.

To provide a convenient and simple parameter to express the intensity
of the shock heating we define a characteristic time for disruption
(see e.g. Spitzer 1987).  The characteristic time is related to the
number of periastron passages a satellite needs to reach the zero
energy state and become unbound:
\begin{equation}
t_{\rm dis}=T \cdot {E_0 \over \langle \Delta E_{\rm hm}\rangle_{\rm act}}
 \;, 
\end{equation}
where $E_0$ is the binding energy of the already tidally truncated
satellite evaluated at the half mass radius $r_{\rm hm}$: $ E_0=0.25 G
M_{\rm s,tid}/R_{\rm hm}$ (Gnedin, Lee \& Ostriker 1999), (for all the
detail of the calculation see Gnedin, Herquist \& Ostriker 1999 and
Taffoni et al 2001).

The rate of mass loss by the satellite due to tidal shocks can be
written in terms of the characteristic disruption time as:
\begin{equation}
-{1 \over M } {d\,M \over d \,t} \propto {1 \over t_{\rm dis}} \;,
\end{equation}
(Spitzer 1987, pp. 115-117).
 
\begin{figure}
\plottwo{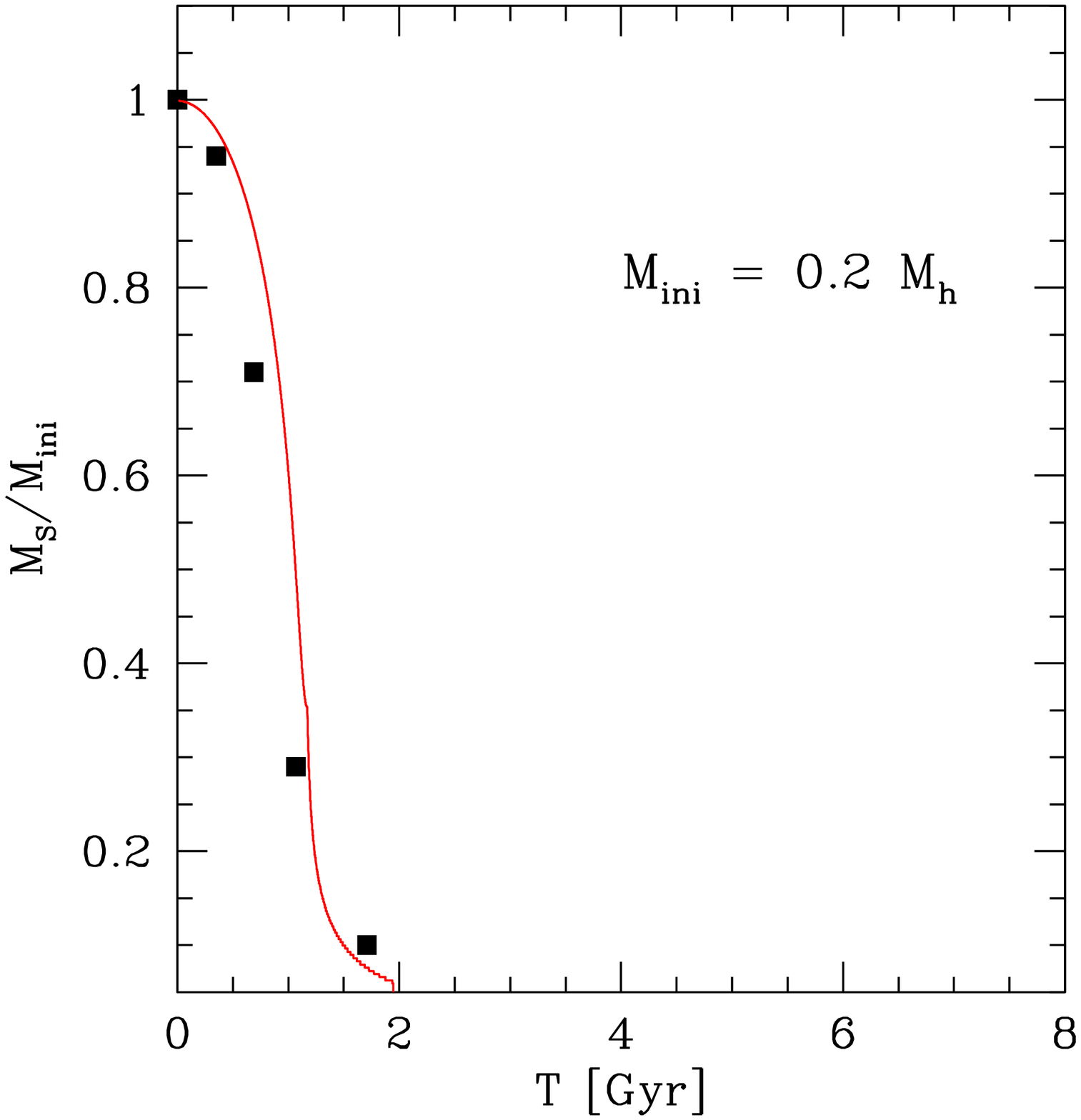}{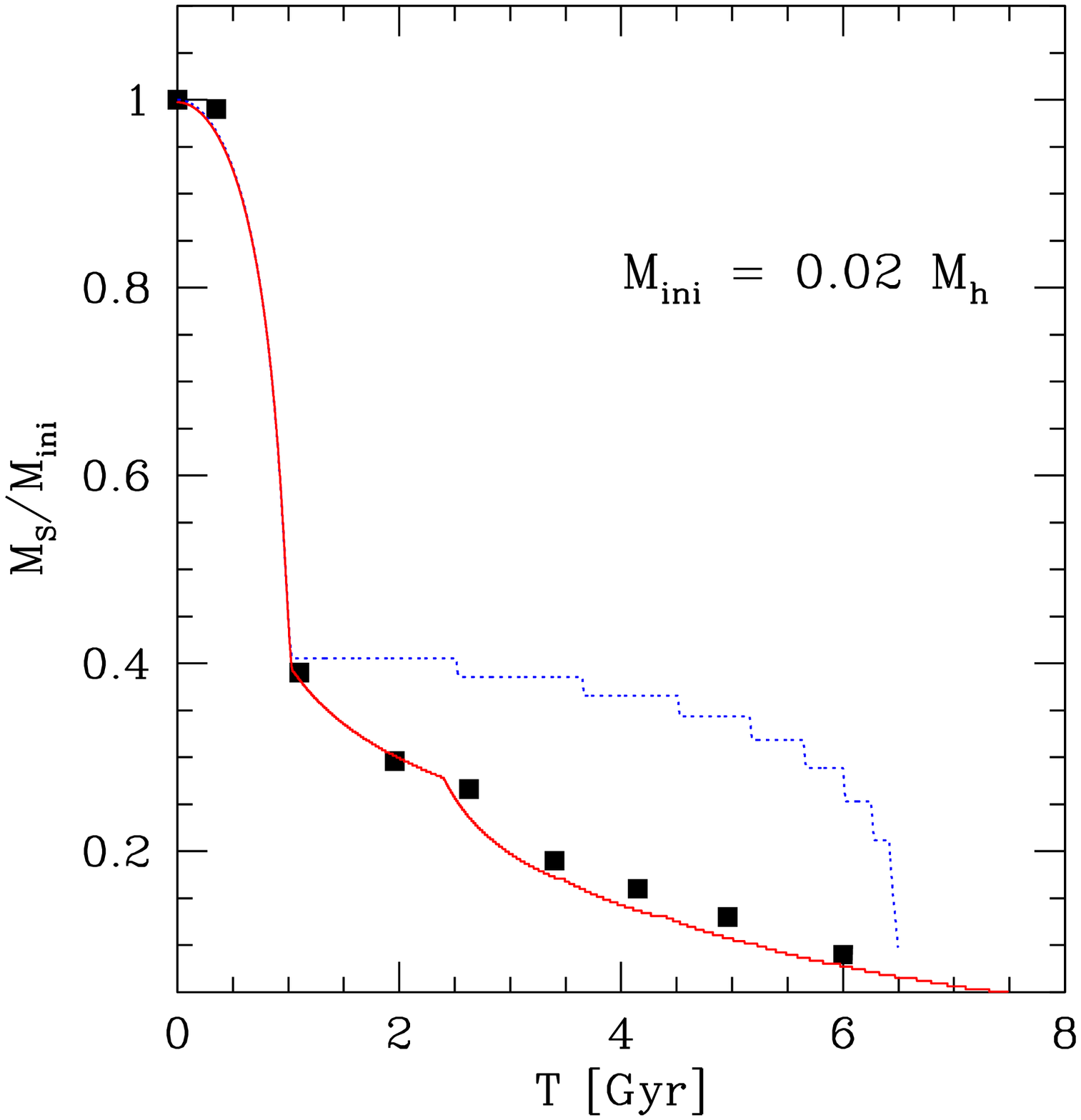}
\caption{The instantaneous mass  of a satellite  
  for initial orbital parameters $\epsilon=0.6$ and $r_{\rm c}(E)=0.5$
  and for an initial apoastron $r=0.74$ R$_{\rm vir}$, where R$_{\rm
    vir}$ is the primary halo virial radius.  The left plot shows the
  instantaneous mass normalized to the initial one for a satellite of
  initial mass $1/5$ of the primary halo mass, the solid line is the
  analytical model and the points are the simulation data.  The right
  plot shows the same quantity for a satellite of initial mass $1/50$
  the mass of the primary halo. The dotted line is the mass-loss for a
  satellite that experiences only the tidal truncation at pericenter.}
\end{figure}

\subsection{Tidal stripping with dynamical friction}  
The dynamical friction force is related to the mass and the size of
the satellite, and it is highly affected by the mass loss induced by
the tidal perturbation.  While the satellite reduces its mass the drag
force becomes less intense and the dynamical friction time increases.
On the other hand the typical energy of the shock increases because
the orbit shrinks due to dynamical friction.  The two processes are
thus strongly connected and cooperate to dissolve the satellite.

We consider a simple model in which a spherical satellite halo is
orbiting inside a spherical primary halo, both being described by a
NFW profile, with concentration $c_{\rm s} = 30$ and $c_{\rm h}=15$,
respectively.  We follow the dynamical evolution of satellites on an
orbit with initial orbital parameters $\epsilon=0.6$ and $r_{\rm
  c}(E)=0.5$.  This particular combination of parameters is the most
likley in a cold dark matter Universe (see e.g.  Tormen 1997; Ghigna
et. al 1998).

In figure 1, we show the results for two satellites, a light one with
initial mass $M_{\rm ini}=0.02 M_{\rm h}$, where $M_{\rm h}$ is the
main halo mass, and a heavy one with $M_{\rm ini}=0.2 M_{\rm h}$.
During the first periastron passage both satellites are tidally
truncated and reduce their mass of about 60 \%. The amount of mass
stripped by the background during this first phase of evolution
depends on the initial orbital parameters and on the density profile
of both the main halo and the satellite.  Then, at each periastron
passage, satellites are tidally shocked and they loose mass with a
rate determined by eq. 5.

We compare the results obtained within our model with an N-body
simulation performed with PKDGRAV, a high-performance parallel binary
treecode developed by the HPCC group in Seattle (Dikaiakos \& Stadel
1996; Stadel, Wadsley \& Quinn, in preparation).  PKDGRAV has
multistepping capabilities which makes it ideal for following
accurately and efficiently a rapidly varying density field like that
typical of simulations with tidal interactions (see Mayer et al.
2001).  We used 20.000 particles for the satellite and 50.000
particles for the primary halo.  We note that a complete description
that accounts for both the tidal truncation and the evaporation due to
tidal shocks can reproduce with good accuracy the mass loss of the
N-Body satellite; a simple tidal truncation is instead insufficient
(see. figure 1, right plot).

\section{Conclusions}
The simple model presented allows to investigate the fate of
substructures inside a dark matter halo.  It is importnat to know if a
satellite merges with the central object of the main system or it
evaporates without interacting with any other halo; infact, the impact
on the formation/evolution of host galaxies in both the primary and
satellite haloes will be radically different in one or the other case.

We note that the dynamical evolution of a substructure is the result
of the interplay between dynamical friction and tidal evaporation. We
emphasize that a simple analytical model which does not include the
tidal shocks cannot describe the dynamical evolution of the dark
matter satellites properly.

The efficiency of the tidal disruption depends on the initial orbital
parameters and on the density profile of both the primary halo and the
satellite.  Highly concentrated profiles induce a more intense tidal
shock on the satellite.  On the other hand, highly concentrated
satellites will be more stable to evaporation, responding more
adiabatically to the tidal field.

We explored a wide parameter space within our analytical model and ran
several N-Body simulations for comparison.

We finally identified three different regimes in which the evolution
of the satellites can take place:
\begin{enumerate}
\item High mass satellites ($M_{\rm s} > 0.2 M_{\rm h}$) reduce their
  mass but the efficiency of the dynamical friction is high enough to
  drive them to the center as if they were rigid.  {\it Merging} is
  thus the fate of such satellites.
\item For low mass satellites $M_{\rm s} << 0.01 M_{\rm h}$, the
  dynamical friction time is longer than the Hubble time and the tidal
  evaporation will eventually disrupt such haloes on cosmological
  orbits.  {\it Distruption} or {\it Survival} is thus the fate of
  such satellites depending on the value of their central density and
  on their average orbital radius.
\item For $0.01 M_{\rm h} < M_{\rm s} < 0.1 M_{\rm h}$ both the
  dynamical friction and tidal disruption contribute to the death of
  substructure.  Satellites can decay substantially towards the center
  and merge, but only after a conspicious mass loss.
\end{enumerate}

\section*{Acknowledgements}

We thank Tom Quinn and Joachim Stadel for providing us with the N-Body
code PKDGRAV.
G.Taffoni is grateful to Pierluigi Monaco and James Taylor for useful
discussions.

\end{document}